\pdfoutput=1
\RequirePackage{ifpdf}
\ifpdf 
\documentclass[pdftex]{sigma}
\else
\documentclass{sigma}
\fi

\numberwithin{equation}{section}

\newtheorem{Theorem}{Theorem}[section]

 { \theoremstyle{definition}

\newtheorem{Remark}[Theorem]{Remark} }

\def\te{\text{e}}
\def\td{\text{d}}
\def\bt{\triangleright}
\def\ct{\blacktriangleright}

\begin{document}
\allowdisplaybreaks

\newcommand{\arXivNumber}{2112.12038}

\renewcommand{\PaperNumber}{022}

\FirstPageHeading

\ShortArticleName{Deformed Quantum Phase Spaces, Realizations, Star Products and Twists}

\ArticleName{Deformed Quantum Phase Spaces, Realizations,\\ Star Products and Twists}

\Author{Stjepan MELJANAC~$^{\rm a}$ and Rina \v{S}TRAJN~$^{\rm b}$}

\AuthorNameForHeading{S.~Meljanac and R.~\v{S}trajn}

\Address{$^{\rm a)}$~Division of Theoretical Physics, Ruder Bo\v skovi\'c Institute,\\
\hphantom{$^{\rm a)}$}~Bijeni\v{c}ka cesta 54, 10002 Zagreb, Croatia}
\EmailD{\href{mailto:meljanac@irb.hr}{meljanac@irb.hr}}

\Address{$^{\rm b)}$~Department of Electrical Engineering and Computing, University of Dubrovnik,\\
\hphantom{$^{\rm b)}$}~\'Cira Cari\'ca 4, 20000 Dubrovnik, Croatia}
\EmailD{\href{mailto:rina.strajn@unidu.hr}{rina.strajn@unidu.hr}}

\ArticleDates{Received December 27, 2021, in final form March 14, 2022; Published online March 23, 2022}

\Abstract{We review deformed quantum phase spaces and their realizations in terms of undeformed phase space. In particular, methods of calculation for the star product, coproduct of momenta and twist from realizations are presented, as well as their properties and the relations between them. Lie deformed quantum phase spaces and Snyder type spaces are considered. Examples of linear realizations of the $\kappa$-Minkowski spacetime are elaborated. Finally, some new results on quadratic deformations of quantum phase spaces and a generalization of Yang and triply special relativity models are presented.}

\Keywords{deformed quantum phase spaces; realizations; star products; twists}

\Classification{81R60}

\section{Introduction}

Noncommutative NC spaces appeared in theoretical physics in the efforts to understand and model Planck scale phenomena. The first proposed model of NC geometry was that of the Snyder spacetime \cite{Snyder}. In order to obtain quantum gravity models that reconcile general relativity (GR) and quantum mechanics (QM), one of the main ideas is to introduce noncommutative quantum spacetimes \cite{dfr1,dfr2,garay} and also look for quantum deformations of the quantum-mechanical relativistic phase space algebra \cite{yang}. For example, a widely studied model is the $\kappa$-Minkowski spacetime, where the parameter $\kappa$ is usually interpreted as the Planck mass or the quantum gravity scale and the coordinates themselves close a Lie algebra \cite{gacluknow,dasklukwor,freikowgliknow, luknowru,luknowruto,majidru}.

A successful approach to noncommutative geometry is based on the formalism of Hopf algebras \cite{chapre, majid}, describing the relativistic symmetries of the quantum spacetime \cite{arzkowglik, ascdimkuliwess}. The $\kappa$-Poincar\'e quantum group \cite{gillkosmajmasku, maslanka,zakrzewski}, as a possible quantum symmetry of the $\kappa$-Minkowski spacetime, allows for the study of deformed relativistic spacetime symmetries and the corresponding dispersion relations \cite{gac,gacmajid,ascborpach2,ascborpach,meljmigs}. It is an example of a Hopf algebra, where the algebra sector is the same as that of the Poincar\'e algebra, but the coalgebra sector is deformed. In general, in the Hopf algebra framework, it is possible to deform the Hopf algebra using a twist element which satisfies the 2-cocycle condition, which again produces a Hopf algebra, with the algebra sector unaltered and the coalgebra sector deformed. Deformations of relativistic symmetries play an important role in the study of phenomenologically relevant effects of quantum gravity \cite{Addazi, gac1,gac2,gacsmolstar,kowgliknow,kowgliksmol,magsmol}. The interplay between spacetime curvature, speed of light and quantum deformations of relativistic symmetries was presented in~\cite{ballgubmer}.

A powerful tool in the study of NC spaces is that of realizations in terms of the Weyl--Heisenberg algebra \cite{battisti1,beckers,durmeljsamst,
govgupharmelj,govgupharmelj2,
hallsz,meljmigs,meljpikgup,meljk-j,meljk-jst,meljsamstgup,meljst,dong}. Namely, the NC coordinates are expressed in terms of the commutative coordinates and the corresponding momenta, which allows one to simplify the methods of calculation on the deformed spacetime. Every realization corresponds to a specific ordering and one special example is the Weyl realization, related to the symmetric ordering. Exponential formulas \cite{battisti2,meljsamst,meljsamst1,meljsamst2,meljsksvrt} are related to the deformed coproduct of momenta in NC spaces and also appear in the computations of star products, which are both needed for the definition of a field theory, the notion of differential calculus and other calculations in a NC spacetime \cite{gacarz,chryok,grac-bonlizz,jurmeljpiks,madore}. Exponential formulae, which were used to obtain the coproducts and star products were also presented in \cite{bormeljpach,meljpachpik,meljs}. A Lie deformed phase space obtained from a twists in the Hopf algebroid approach was considered in~\cite{jurkovmelj,jurmeljs,lukmeljpikwor,lukmeljwor}.

A class of deformed quantum phase spaces, i.e., a deformed Heisenberg algebra, is generated with NC coordinates $\hat{x}_\mu$ and commutative momenta $p_\mu$ defined as
\begin{equation}\label{prva}
[\hat{x}_\mu, \hat{x}_\nu] = {\rm i} \Theta_{\mu\nu} (l, \hat{x}, p), \qquad [p_\mu,\hat{x}_\nu]=-{\rm i}\varphi_{\mu\nu}(l,p),\qquad [p_\mu,p_\nu]=0,
\end{equation}
with all Jacobi identities satisfied and $l$ a real parameter of order of the Planck length. There are also deformed quantum phase space models with NC momenta $\hat{p}_\mu$ including the cosmological radius $R$ \cite{kowgliksmol, yang}. The undeformed phase space is defined as
\begin{gather*}
[x_\mu, x_\nu] = 0,\qquad
[p_\mu, p_\nu] = 0,\qquad
[x_\mu, p_\nu] = -{\rm i} \eta_{\mu\nu}, \qquad \mu, \nu = -1,\dots ,n-1,
\end{gather*}
where $\eta_{\mu\nu}=\operatorname{diag} (-1,1,\dots ,1)$, $x_\mu$ are coordinates and $p_\mu$ are momenta.

Examples of NC spaces where $\Theta_{\mu\nu}$ \eqref{prva} do not depend on the momenta $p_\mu$ are the canonical theta space \cite{dfr2,nairpoli}, with $\Theta_{\mu \nu} = {\rm const}$, Lie algebra type spaces \cite{gacarz,gacluknow,batmajid,chryok,dasklukwor,daskwal,freikowgliknow,freiliv3d,gnatenko,lukmeljpikwor,luknowru,luknowruto,lukwor,majidru,meljpikgup,
miwayu}, for which $\Theta_{\mu\nu} = {\rm i} C_{\mu\nu\lambda} \hat{x}_\lambda$, and quadratic deformations of Minkowski space, with $\Theta_{\mu \nu} = \Theta^{\mu \nu \rho \sigma} \hat{x}_\rho \hat{x}_\sigma$ \cite{lukwor, Wess}.

In the Snyder space \cite{bankulsam,bankumroy,carrmig,girliv,guohuangwu,lustern2,lustern1,meljmigs2,mig1,mig4,mig2,mig3,
mig,migs,migs2,romzam1,romzam2,Snyder}, one has $[\hat{x}_\mu, \hat{x}_\nu] = {\rm i} l^2 M_{\mu \nu}$, where $M_{\mu \nu}$ are Lorentz generators, NC coordinates $\hat{x}_\mu$ do not close an algebra between themselves, but $\hat{x}_\mu$ and $M_{\mu \nu}$ close a Lie algebra. The Snyder space and the $\kappa$ deformed Snyder space \cite{meljsamst1,meljsamst2} lead to a non associative star product and non coassociative coproduct \cite{battisti2,girliv}. Recently, the extended Snyder model and the $\kappa$ deformed extended Snyder model with additional tensorial coordinates were proposed \cite{meljmig1,meljmig3,meljmig2}. These models are of Lie algebra type NC spaces in which star products are associative and coproducts are coassociative~\cite{meljmig3}.

Some new developements in the applications of NC geometry to physics can be found in \cite{carmcortrel2,carmcortrel3,carmcortrel1,kupkurvit,kupvit,
lizzvit,matwall2,matwall1,relancio1,relancio2}.

In this review we survey the above mentioned types of deformed quantum phase spaces, their properties and the relations between realizations of NC coordinates in terms of undeformed phase space, star products and twists. Many technical results important for this review have appeared previously in the literature and are cited appropriately. However this list of references is not exhaustive. We also present some new results in Sections~\ref{section3},~\ref{section4},~\ref{section6} and~\ref{section7}.

The plan of paper is as follows. In Section~\ref{section2}, the Lie deformed quantum phase space and realizations are presented. In Section~\ref{section3}, a star product from realizations is constructed. In Section~\ref{section3.1}, we present the Snyder space and its extension with tensorial coordinates. In Section~\ref{section4}, the coproduct of momenta and twist from star product and realizations are obtained. In Section~\ref{section5}, some examples of linear realizations of the $\kappa$ deformed Minkowski spacetime, specially the right covariant, left covariant and light like realizations, are revisited. In Section~\ref{section6}, some aspects of quadratic deformations of quantum phase spaces are elaborated, specially in Section~\ref{section6.1}, quadratic deformations of the Minkowski space from dilatation. A generalization of Yang and triply special relativity models is given in Section~\ref{section7}.

\section{Lie deformed quantum phase space and realizations}\label{section2}

The undeformed quantum phase space is defined with coordinates $x_\mu$ and momenta $p_\mu$
\begin{gather*}
[x_\mu,x_\nu]=0, \quad [p_\mu,p_\nu]=0, \qquad [p_\mu,x_\nu]=-{\rm i}\eta_{\mu\nu},\qquad \mu,\nu =0,1,\dots ,n-1,
\end{gather*}
where $\eta_{\mu\nu}=\operatorname{diag}(-1,1,\dots ,1)$. The generalization of $\eta_{\mu\nu}$ to a metric with arbitrary Lorentz signature is straightforward.

We consider the deformed quantum phase space defined with noncommutative (NC) coordinates $\hat{x}_\mu$ and momenta $p_\mu$ of the type
\begin{gather}\label{defWH}
[\hat{x}_\mu,\hat{x}_\nu]={\rm i}\hat{x}_\alpha C_{\mu\nu\alpha}(l,p)+ {\rm i} d_{\mu\nu}(l,p), \qquad [p_\mu,\hat{x}_\nu]=-{\rm i}\varphi_{\mu\nu}(l, p),\qquad [p_\mu,p_\nu]=0,
\end{gather}
where $l$ is a parameter of order of the Planck length and summation over repeated indices is assumed. For $l =0$, $C_{\mu\nu\alpha}(0,p) = 0$ and $d_{\mu\nu}(0,p) = 0$. Functions $C_{\mu\nu\alpha}$ and $\varphi_{\mu\nu}$ depend on momenta~$p_\alpha$, where $C_{\mu\nu\alpha}$ are a generalization of the structure constants.

For example, the original Snyder space is defined with
\begin{gather}\label{originalSnyder}
[\hat{x}_\mu,\hat{x}_\nu]={\rm i}l^2M_{\mu\nu},
\end{gather}
where $M_{\mu\nu}$ are Lorentz generators. The realization Snyder proposed~\cite{Snyder} is given by
\begin{gather*}
\hat{x}_\mu=x_\mu +l^2(x\cdot p)p_\mu,\qquad M_{\mu\nu}=x_\mu p_\nu -x_\nu p_\mu =\hat{x}_\mu p_\nu -\hat{x}_\nu p_\mu.
\end{gather*}
Hence
\begin{gather*}
C_{\mu\nu\alpha}(l,p)=l^2(\eta_{\mu\alpha}p_\nu -\eta_{\nu\alpha}p_\mu).
\end{gather*}

If $C_{\mu\nu\lambda}$ are structure constants, then NC coordinates $\hat{x}_\mu$ close a Lie algebra. A perturbative construction of $\varphi_{\mu\nu}(l,p)$ corresponding to the symmetric ordering can be found in~\cite{durmeljsamst}. From $[p_\mu,\hat{x}_\nu]=-{\rm i}\varphi_{\mu\nu}(l,p)$,~\eqref{defWH}, it follows that the realization of NC coordinates $\hat{x}_\mu$ can be written as
\begin{gather}\label{hatx}
\hat{x}_\mu=x_\alpha \varphi_{\alpha\mu}(l, p)+ \chi_\mu (l, p),
\end{gather}
and the inverse is given by
\begin{gather*}
x_\nu =(\hat{x}_\mu -\chi_\mu (l,p))\big(\varphi^{-1}\big)_{\mu\nu},
\end{gather*}
and
\begin{gather}\label{pmu}
p_\mu=-{\rm i}\frac{\partial}{\partial x_\mu}.
\end{gather}
If $l=0$, $\varphi_{\mu\nu}(0,p) = \eta_{\mu\nu}$. All Jacobi relations for the class of deformed quantum phase spaces/deformed Weyl--Heisenberg algebras~\eqref{defWH} (obtained from~\eqref{hatx}) are satisfied.

Realizations of the type given in equation~\eqref{hatx}, used in physical applications, were studied for example for $\kappa$-Minkowski spaces in \cite{govgupharmelj,govgupharmelj2,jurmeljpiks,meljmigs,meljpikgup,meljk-jst,meljsamstgup,
meljst}, for Snyder spaces in \cite{battisti1,battisti2,meljmigs2}, and for the extended Snyder model with tensorial coordinates in \cite{meljmig1,meljmig3,meljmig2}.

A special class of deformed quantum phase spaces/deformed Weyl--Heisenberg algebras for which $\varphi_{\mu\nu}(l,p)$ is at most linear in $p_\alpha$ is given by
\begin{gather*}
\hat{x}_\mu =x_\mu +l K_{\beta\mu\alpha} x_\alpha p_\beta + \chi_\mu (l, p), \qquad K_{\beta \mu\alpha} \in \mathbb{R},
\end{gather*}
where
\begin{gather*}
\varphi_{\alpha\mu}(l, p)= \eta_{\alpha\mu} +l K_{\beta\mu\alpha} p_\beta.
\end{gather*}
For $\chi_\mu(l, p)=0$, it holds
\begin{gather*}
[\hat{x}_\mu,\hat{x}_\nu] ={\rm i}l(K_{\mu\nu\alpha}-K_{\nu\mu\alpha}) x_\alpha +{\rm i}l^2 (K_{\beta\mu\alpha} K_{\alpha\nu\sigma} -K_{\beta\nu\alpha} K_{\alpha\mu\sigma}) x_\sigma p_\beta.
\end{gather*}
For example, the linear realization for the case of the $\kappa$ Snyder deformation was presented in~\cite{meljsamst2}.

The Lie algebra is closed in NC coordinates $\hat{x}_\mu$ if
\begin{gather*}
K_{\beta\mu\lambda} K_{\lambda\nu\alpha} -K_{\beta\nu\lambda} K_{\lambda\mu\alpha} = (K_{\mu\nu\lambda} - K_{\nu\mu\lambda}) K_{\beta\lambda \alpha},
\end{gather*}
and the structure constants are
\begin{gather*}
C_{\mu\nu\alpha} =K_{\mu\nu\alpha} -K_{\nu\mu\alpha},
\end{gather*}
see for example \cite{jurmeljpik,lukmeljpikwor,lukmeljwor,meljpik,meljpikgup,meljskos}.

\section{Star product from realizations}\label{section3}

Let us define the $\bt$ action
\begin{gather*}
x_\mu \bt f(x) = x_\mu f(x),\qquad
p_\mu \bt f(x) = -{\rm i}\frac{\partial f(x)}{\partial x_\mu}.
\end{gather*}
For the realization given in equation~\eqref{hatx}, using $\bt$, it follows
\begin{gather*}
 p_\mu \bt \te ^{{\rm i}qx} =q_\mu \te ^{{\rm i}qx}, \qquad
\te ^{{\rm i}k_\alpha \hat{x}_\alpha} \bt \te ^{{\rm i}q_\beta x_\beta} = \te ^{{\rm i}J_\alpha (k,l,q)x_\alpha +{\rm i}h(k,l,q)}, \qquad k_\alpha,q_\alpha \in M_n,
\end{gather*}
where $\hat{x}_\mu=x_\alpha \varphi_{\alpha\mu}(l,p) +\chi_\mu(l,p)$ and $M_n$ denotes the Minkowski space, for some~$J$,~$h$~\cite{meljs}. If $k_\alpha=0$, $J_\mu(0,l,q)=q_\mu$, $h(0,l,q)=0$. If $q_\alpha=0$, $J_\mu(k,l,0)=K_\mu(k,l)$. If $l=0$, $J_\mu(k,0,q)=k_\mu+q_\mu$. $J_\mu(k,l,q)$ and $h(k,l,q)$ can be constructed perturbatively using \cite[Theo\-rems~1 and~2]{meljs}.

Compact results for $J_\mu(tk,l,p)$ and $h(tk,l,p)$, where $p_\alpha$ is the momentum operator~\eqref{pmu}, are given by
\begin{gather*}
 J_\mu(tk,l,p)= \big(\te ^{tk_\alpha O_\alpha}\big)(p_\mu),\qquad
 h(tk,l,p)= \left(\frac{\te ^{tO}-1}{O}\right)(k_\beta \chi_\beta(l,p)),
\end{gather*}
where
\begin{gather*}
O=k_\alpha O_\alpha, \qquad O_\alpha=\operatorname{ad}_{-{\rm i}x_\beta\varphi_{\beta\alpha}(l,p)}.
\end{gather*}
$J_\mu(tk,l,p)$ and $h(tk,l,p)$ are unique solutions of the partial differential equations
\begin{gather}
\frac{\partial J_\mu(tk,l,p)}{\partial t} = k_\beta \varphi_{\mu\beta} (J(tk,l,p)),\qquad
\frac{\partial h(tk,l,p)}{\partial t} = k_\beta \chi_\beta(J(tk,l,p)), \label{difjedzaJ}
\end{gather}
with boundary conditions
\begin{gather*}
J_\mu (0,l,p) = p_\mu,\qquad
h(0,l,p)=0.
\end{gather*}
From $J_\mu(k,l,q)$ and $h(k,l,q)$ we can obtain the star product
\begin{gather}\label{starprod}
\te ^{{\rm i}kx}*\te ^{{\rm i}qx} =\te ^{{\rm i}K^{-1}(k)\hat{x} -{\rm i}h(K^{-1}(k),0)} \bt \te ^{{\rm i}qx} =\te ^{{\rm i}x\mathcal{D}(k,l,q)+{\rm i}\mathcal{G}(k,l,q)},
\end{gather}
where
\begin{gather}
\mathcal{D}_\mu (k,l,q)= J_\mu\big(K^{-1}_\mu(k,l),l,q\big)=\left( \exp\left( K_\beta^{-1}(k,l) \varphi_{\alpha\beta}(l,q)\frac{\partial}{\partial q_\alpha}\right)\right)(q_\mu),\label{Dmu}\\
\mathcal{G}(k,l,q) =h\big(K^{-1}(k,l),l,q\big)-h\big(K^{-1}(k,l),l,0\big).\label{Gmu}
\end{gather}
$K^{-1}$ is the inverse map
\begin{gather*}
K^{-1}_\mu(K(k,l)) =K_\mu\big(K^{-1}(k,l)\big) =k_\mu.
\end{gather*}
The star product, \eqref{starprod}, and the method of calculation are a generalization of the method first proposed in \cite{meljsamst,meljsamst1}. The method from \cite{meljsamst,meljsamst1} was applied in \cite{battisti2,bormeljpach,meljpik,meljmercpik,meljmigs2,meljpachpik,meljsamst2,
meljmig1,meljmig3,meljmig2,meljsksvrt,meljskos}. The star product, \eqref{starprod}, can be associative or nonassociative. For example, the star product for the Snyder model is nonassociative~\cite{battisti2,girliv}. If NC coordinates~$\hat{x}_\mu$,~\eqref{hatx}, close a Lie algebra, then the corresponding star product is associative~\cite{meljk-j}.

According to the PBW theorem, if coordinates $\hat{x}_\mu$, $\mu=0,1,\dots ,n-1$ generate a Lie algebra~$\hat{g}$ and $x_\mu =\hat{x}_\mu \bt 1$, $\mu=0,1,\dots ,n-1$ generate a commutative algebra~$g$, then
\begin{itemize}\itemsep=0pt
\item[i)] enveloping algebras $\mathcal{U}(\hat{g})$ and $\mathcal{U}(g)$ are isomorphic,
\item[ii)] if $\hat{f}\bt 1=f$, $U(\hat{g}) \bt 1 =U(g)$, then we define the inverse map $\ct$
\begin{gather*}
f\ct 1=\hat{f}, \qquad g\ct 1=\hat{g},
\end{gather*}
and the star product
\begin{gather*}
 f*g=\hat{f}\hat{g}\bt 1,\qquad
 (f*g)\ct 1 =\hat{f}\hat{g},
\end{gather*}
where $\hat{f},\hat{g}\in \mathcal{U}(\hat{g})$ and $f,g\in \mathcal{U}(g)$. This star product is associative
\begin{gather*}
f*(g*h)=(f*g)*h, \qquad f,g,h\in \mathcal{U}(g).
\end{gather*}
\end{itemize}
Namely
\begin{gather*}
f*(g*h)\ct 1=\hat{f}\big(\hat{g}\hat{h}\big)=\hat{f}\hat{g}\hat{h},\qquad
(f*g)*h\ct 1=\big(\hat{f}\hat{g}\big)\hat{h}=\hat{f}\hat{g}\hat{h}.
\end{gather*}

Generally, if the star product, \eqref{starprod}, is associative, i.e.,
\[
\big(\te ^{{\rm i}k_1x}* \te ^{{\rm i}k_2 x}\big)* \te ^{{\rm i}k_3x} = \te ^{{\rm i}k_1x}*\big( \te ^{{\rm i}k_2x}* \te ^{{\rm i}k_3x}\big),
\] then it holds
\begin{gather}\label{zahtjevasoc}
\mathcal{D}_\mu(\mathcal{D}(k_1,l,k_2),l,k_3) =\mathcal{D}_\mu (k_1,l,\mathcal{D}(k_2,l,k_3)),
\end{gather}
and
\begin{gather*}
\mathcal{G}(k_1,l,k_2) +\mathcal{G}(\mathcal{D}(k_1,l,k_2),l,k_3) =\mathcal{G} (k_2,l,k_3) +\mathcal{G}(k_1,l,\mathcal{D}(k_2,l,k_3)).
\end{gather*}

Addition of momenta is defined with
\begin{gather*}
k_\mu \oplus q_\mu=\mathcal{D}_\mu(k,l,q).
\end{gather*}
Note that $\mathcal{D}_\mu(k,l,q)\in \mathbb{R}$.

For example, let the NC coordinates $\hat{x}_i$, $i=1,2,3$ close the $\mathfrak{su}(2)$ algebra
\begin{gather*}
[\hat{x}_i,\hat{x}_j]=2{\rm i}l\epsilon_{ijk}\hat{x}_k,
\end{gather*}
and the realization of $\hat{x}_i$ is given by
\begin{gather*}
\hat{x}_i =x_i\sqrt{1-l^2p^2}+l\epsilon_{ijk} x_jp_k, \qquad p^2=p_1^2+p_2^2+p_3^2.
\end{gather*}
Then
\begin{gather*}
[\hat{x}_i,p_j]= {\rm i}\big(\delta_{ij} \sqrt{1-l^2p^2}+l\epsilon_{ijk}p_k\big),
\end{gather*}
and
\begin{gather*}
k_i\oplus q_i =\mathcal{D}_i(k,q) =k_i\sqrt{1-l^2q^2}+ \sqrt{1-l^2k^2}q_i +l\epsilon_{ijk}k_j q_k.
\end{gather*}
This example of the $\mathfrak{su}(2)$ NC space appeared in connection with the 3 dimensional quantum gravity model \cite{freiliv3d}, see also \cite{kupvit2015, meljsksvrt}.

Note that the universal formula for Lie algebra generators as formal power series of the corresponding structure constants with coefficients in Bernoulli numbers was given in \cite{durmeljsamst,gutt}. The explicit star product of the cotangent bundle of a Lie group~\cite{gutt} corresponds to the star product \eqref{starprod}, where $\hat{x}$ is expressed in terms of the universal formula, i.e., $\varphi(l,p)$ is the generating function for Bernoulli numbers, related to the symmetric ordering~\cite{durmeljsamst}.

\subsection{Snyder space and an extension with tensorial coordinates}\label{section3.1}

Examples of non associative star products are related to the following realizations of coordinates~$\hat{x}_\mu$
\begin{gather*}
\hat{x}_\mu = x_\mu \varphi_1 \big(l^2 p^2\big) +l^2(x\cdot p)p_\mu \varphi_2\big(l^2 p^2\big).
\end{gather*}
Then the commutation relations $[\hat{x}_\mu,\hat{x}_\nu]$ lead to a generalized Snyder algebra~\cite{meljmigs2}
\begin{gather*}
[\hat{x}_\mu,\hat{x}_\nu]= {\rm i}l^2 M_{\mu\nu} \psi \big(l^2 p^2\big),
\end{gather*}
where $M_{\mu\nu}=x_\mu p_\nu-x_\nu p_\mu$ are Lorentz generators. Specially, for $\psi\big(l^2 p^2\big)=1$, it becomes the Snyder algebra originally proposed in~\cite{Snyder}.

In the original Snyder model NC coordinates $\hat{x}_\mu$ do not close a Lie algebra and the corresponding star products are non associative. For example, the star product corresponding to the realization
\begin{gather}\label{Snrealiz}
\hat{x}_\mu =x_\mu +l^2 (x\cdot p)p_\mu
\end{gather}
leads to
\begin{gather*}
x_\mu *x_\nu =x_\mu x_\nu,
\end{gather*}
and
\begin{gather*}
x_\mu *(x_\nu *x_\rho) = x_\mu x_\nu x_\rho -l^2 (\eta_{\mu \nu} x_\rho +\eta_{\mu \rho} x_\nu), \\
(x_\mu *x_\nu) *x_\rho = x_\mu x_\nu x_\rho -\frac{l^2}{2} (\eta_{\mu \rho} x_\nu +\eta_{\nu \rho} x_\mu +2\eta_{\mu \nu}x_\rho).
\end{gather*}
Hence, the star product is non associative
\begin{gather*}
(x_\mu * x_\nu) *x_\rho -x_\mu *(x_\nu *x_\rho) =\frac{l^2}{2}(\eta_{\mu\rho}x_\nu -\eta_{\nu\rho} x_\mu).
\end{gather*}

Using the realization in equation~\eqref{Snrealiz}, and equations~\eqref{difjedzaJ},~\eqref{Dmu}, the function $\mathcal{D}_\mu (k,l,q)$, defining the star product~\eqref{starprod}, was obtained in~\cite{battisti2}
\begin{gather*}
\mathcal{D}_\mu (k,l,q) =\frac{1}{1-l^2(k\cdot q)} \left( k_\mu -\frac{l^2}{1+\sqrt{1+l^2k^2}}k_\mu (k\cdot q) +\sqrt{1+l^2k^2}q_\mu \right),
\end{gather*}
for which equation~\eqref{zahtjevasoc} is not satisfied, implying that the star product is non associative. The corresponding coproduct
\begin{gather*}
\Delta p_\mu =\frac{1}{1-l^2p_\alpha \otimes p^\alpha} \left( p_\mu \otimes 1 -\frac{l^2}{1+\sqrt{1+l^2p^2}} p_\mu p_\alpha \otimes p^\alpha +\sqrt{1+l^2p^2} \otimes p_\mu \right)
\end{gather*}
is non coassociative.

Note that in the original Snyder space NC coordinates $\hat{x}_\mu$ and Lorentz generators $M_{\mu\nu}$ close a Lie algebra, but $M_{\mu\nu}\bt 1=0$. Since NC coordinates $\hat{x}_\mu$ do not close a Lie algebra between themselves, additional tensorial coordinates $\hat{x}_{\mu\nu}$ are introduced \cite{girliv,meljmig1} instead of $M_{\mu\nu}$ and they satisfy
\begin{gather*}
\hat{x}_{\mu\nu} \bt 1=x_{\mu\nu},
\end{gather*}
where $x_{\mu\nu}$ are commutative tensorial coordinates. Both the coordinates~$\hat{x}_{\mu\nu}$ and~$x_{\mu\nu}$ and the canonical momenta~$p_{\mu\nu}$ transform as Lorentz generators~$M_{\mu\nu}$. Consequently, the new star product is associative and the coproduct is coassociative \cite{girliv,meljmig1,meljpach}.

Using the above extended Snyder space, a unification with the $\kappa$-Minkowski space is also proposed in the form of associative realizations of the $\kappa$ deformed extended Snyder model \cite{meljmig3, meljmig2}.

A similar extension with additional commutative tensorial coordinates $\theta_{\mu\nu}$ was proposed in the context of the DFR NC space, changing the constant $\theta_{\mu\nu}$ to tensorial coordinates $\theta_{\mu\nu}$ \cite{3inicijala,amorim}.

\section[Coproduct of momenta and twist from star product and realizations]{Coproduct of momenta and twist from star product\\ and realizations}\label{section4}

The relation between the star product and the twist operator is given by
\begin{gather*}
(f*g)(x)=m\mathcal{F}^{-1}(\bt \otimes 1)(f(x)\otimes g(x)),
\end{gather*}
where $m$ is the multiplication map, $A\otimes B \rightarrow AB$.

Using the star product \eqref{starprod} and the above relation for~$f*g$, a family of twist operators can be writen as \cite{meljmercpik,meljmigs2, meljpachpik}
\begin{gather}\label{twist}
\mathcal{F}^{-1} ={:}\exp ( ({\rm i}(1-u)x_\alpha \otimes 1+u 1\otimes x_\alpha)(\Delta-\Delta_0)p_\alpha ){:}\, \exp( {\rm i}\mathcal{G} (p\otimes 1,1\otimes p)),
\end{gather}
where ${:}\,{:}$ denotes normal ordering, in which the $x$s are to the left of the $p$'s, $u$ is a real parameter and
\begin{gather*}
 \Delta p_\alpha =\mathcal{D}_\alpha (p\otimes 1,1\otimes p),\qquad
 \Delta_0 p_\alpha =p_\alpha \otimes 1+1\otimes p_\alpha,
\end{gather*}
where $\mathcal{D}_\alpha (l,k,q)$ is given in~\eqref{Dmu} and $\mathcal{G}$ in~\eqref{Gmu}. Applying \cite[Theorem~1]{meljs}, we can find the twist in the form without normal ordering, see for example \cite[Section~4]{meljskos}.

It is important to note that $\Delta p_\alpha$ is the coproduct of momenta and it holds
\begin{gather*}
\Delta p_\alpha =\mathcal{F} \Delta_0 p_\alpha \mathcal{F}^{-1}.
\end{gather*}
This can be proved using the identity
\begin{gather*}
\sum_{k,l=0}^{n} (-1)^n \frac{x^kpx^l}{k!l!} =0,\qquad \text{for} \ n\geq 2.
\end{gather*}

Alternatively, we can obtain the twist in the following way. The coproduct $\Delta p_\mu$ is obtained from $\Delta p_\alpha =\mathcal{D}_\alpha (p\otimes 1, 1\otimes p)$ and~\eqref{Dmu}
\begin{gather*}
\Delta p_\mu =\left( \exp\left( K^{-1}_\beta (p) \otimes \varphi_{\alpha\beta} (p) \frac{\partial}{\partial p_\alpha}\right)\right) (1\otimes p_\mu ).
\end{gather*}
The coproduct is coassociative if and only if the star product is associative, i.e., if $\hat{x}_\mu$ close a Lie algebra.

We can define new momenta $p^W_\mu$ corresponding to the Weyl realization
\begin{gather*}
p^W_\mu =K_\mu ^{-1}(p),
\end{gather*}
with the property
\begin{gather*}
p_\mu^W \bt \te ^{{\rm i}K(k)\cdot x} =k_\mu \te ^{{\rm i}K(k)\cdot x},
\end{gather*}
and
\begin{gather*}
\big[p^W_\mu, \te ^{{\rm i}k\cdot \hat{x}}\big]=k_\mu \te ^{{\rm i}k\cdot \hat{x}}.
\end{gather*}
Hence, the coproduct of momenta is
\begin{gather*}
\Delta p_\mu =\te ^{-{\rm i}p^W_\alpha \otimes \hat{x}_\alpha} (1\otimes p_\mu) \te ^{{\rm i}p^W_\beta \otimes \hat{x}_\beta}.
\end{gather*}
Then we obtain the relation between $\Delta p_\mu$ and $\Delta_0 p_\mu$
\begin{gather*}
\Delta p_\mu =\mathcal{F} \Delta_0 p_\mu \mathcal{F}^{-1} =\te ^{-{\rm i}p_\alpha^W \otimes \hat{x}_\alpha} \te ^{{\rm i}p_\beta \otimes x_\beta} (\Delta_0 p_\mu) \te ^{-{\rm i}p_\alpha \otimes x_\alpha} \te ^{{\rm i}p^W_\beta \otimes \hat{x}_\beta}.
\end{gather*}

If $\chi_\mu (l,p)=0$, then the twist operator for the realization $\hat{x}_\mu =x_\alpha \varphi_{\alpha\beta}(l,p)$ is given by
\begin{gather}\label{twist-chije0}
\mathcal{F}^{-1} =\te ^{-{\rm i}p_\alpha \otimes x_\alpha} \te ^{{\rm i}p^W_\beta \otimes x_\gamma \varphi_{\gamma \beta} (l,p)}.
\end{gather}
If $\chi_\mu(l,p)\neq 0$, then the twist $\mathcal{F}^{-1}$ is
\begin{gather}\label{twist2}
\mathcal{F}^{-1} =\te ^{-{\rm i}p_\alpha \otimes x_\alpha} \te ^{{\rm i}p^W_\beta \otimes x_\gamma \varphi_{\gamma \beta}(l,p)} \te ^{{\rm i}\mathcal{G}(p\otimes 1,1\otimes p)},
\end{gather}
and the consistency check is
\begin{gather*}
\hat{x}_\mu =m\mathcal{F}^{-1} (\bt \otimes 1) (x_\mu \otimes 1) = x_\alpha \varphi_{\alpha \mu} (l,p) +\chi_\mu (l,p).
\end{gather*}
Note that the following identity holds
\begin{gather*}
{:} \te ^{{\rm i}(1\otimes x_\alpha)(\Delta-\Delta_0)p_\alpha}{:} = \te ^{-{\rm i}p_\alpha \otimes x_\alpha} \te ^{{\rm i}p_\beta^W \otimes x_\gamma \varphi_{\gamma\beta}(l,p)}.
\end{gather*}

The twists in this section, specially \eqref{twist} and \eqref{twist2}, are given in the Hopf algebroid approach \cite{jurkovmelj,jurmeljs,lukmeljpikwor,lukmeljwor} and generally do not satisfy the cocycle condition in the Hopf algebra sense~\cite{meljsams}. However, in the Hopf algebroid approach, these twists satisfy a generalized cocycle condition if and only if the star product is associative, i.e.,~$\hat{x}_\mu$ close a Lie algebra~\cite{meljsk}. In the next section we give examples for the case when these twists can be transformed into Drinfeld twists satisfying the cocycle condition in the Hopf algebra sense.

Note that another construction of the twist for the original Snyder model \eqref{originalSnyder} was presented in~\cite{meljmigpiks}.

\section{Linear realizations}\label{section5}

For a linear realization
\begin{gather}\label{linrealizhx}
\hat{x}_\mu =x_\mu+ K_{\beta\mu \alpha}x_\alpha p_\beta,
\end{gather}
the differential equation for $J_\mu(tk,q)$ is given by
\begin{gather*}
\frac{\td J_\mu (tk,q)}{\td t}= k_\mu +k_\alpha K_{\beta\alpha\mu}J_\beta (tk,q), \qquad J_\mu(0,q)=q_\mu.
\end{gather*}
The solution for $J_\mu(tk,q)$ is then
\begin{gather}\label{J-lin}
J_\mu(tk,q)= k_\alpha \left( \frac{\te ^{\mathcal{K}(tk)}-\eta}{\mathcal{K}(k)}\right)_{\alpha\mu}+q_\alpha \big( \te ^{\mathcal{K}(tk)}\big)_{\alpha\mu},
\end{gather}
where
\begin{gather}\label{mcK-lin}
\mathcal{K}_{\mu\nu}(k)=K_{\mu\alpha\nu}k_\alpha.
\end{gather}
For $t=1$
\begin{gather*}
J_\mu(k,q)=k_\alpha\left(\frac{\te ^{\mathcal{K}(k)}-\eta}{\mathcal{K}(k)}\right)_{\alpha\mu} +q_\alpha \big(\te ^{\mathcal{K}(k)}\big)_{\alpha \mu}.
\end{gather*}
For $q_\mu=0$
\begin{gather}\label{J-qje0}
J_\mu(k,0)=k_\alpha \left( \frac{\te ^{\mathcal{K}(k)}-\eta}{\mathcal{K}(k)}\right)_{\alpha\mu}=K_\mu(k).
\end{gather}
The inverse of $K_\mu(k)$ is given by
\begin{gather}\label{Kinv-lin}
K^{-1}_\mu(k)= k^W_\mu .
\end{gather}
The expansion of $k_\mu^W$ in terms of $k_\alpha$ was given in the appendix of \cite{meljpik}. Then
\begin{gather}\label{D-lin}
\mathcal{D}_\mu(k,q)= J_\mu \big(K^{-1}(k),q\big)= k_\mu +q_\alpha \big(\te ^{\mathcal{K}(k^W)}\big)_{\alpha\mu},
\end{gather}
and the coproduct for $p_\mu$ is
\begin{gather}\label{deltap-lin}
\Delta p_\mu = p_\mu \otimes 1+ \big(\te ^{\mathcal{K}(p^W)}\big)_{\alpha\mu} \otimes p_\alpha.
\end{gather}

The twist in the Hopf algebroid approach corresponding to the linear realization
$\hat{x}_\mu=x_\mu+K_{\beta\mu\alpha}x_\alpha p_\beta$ is
\begin{gather}\label{twist-lin}
\mathcal{F}=\exp \left( -{\rm i}p_\mu^W \otimes K_{\beta \mu\alpha} x_\alpha p_\beta\right)= \exp \left( -{\rm i}\mathcal{K}_{\beta\alpha}\big(p^W\big) \otimes x_\alpha p_\beta \right),
\end{gather}
where $p_\mu^W=K^{-1}_\mu(p)$.
The coproduct for $p_\mu$, $\Delta p_\mu =\mathcal{F}\Delta_0 p_\mu \mathcal{F}^{-1}$ is identical to the equation above, \eqref{deltap-lin}. Note that
\begin{gather*}
K_\mu (p)=\left( \frac{\te ^{\mathcal{K}(p)}-\eta}{\mathcal{K}(p)}\right)_{\alpha\mu}p_\alpha
\qquad \text{and}\qquad
p_\mu =\left( \frac{\te ^{\mathcal{K}(p^W)}-\eta}{\mathcal{K}\big(p^W\big)}\right)_{\alpha\mu} p_\alpha^W.
\end{gather*}
The expression for $p_\mu^W$ was given in the appendix of~\cite{meljpik}.

For $\kappa$-Minkowski and the corresponding linear realizations an explicit proof of the cocycle codition is given in \cite{jurmeljpik} and more generally in \cite{meljpik}. The conditions under which linear realizations~\eqref{linrealizhx} generate a Lie algebra are given in Section~\ref{section2}. Other examples of linear realizations were presented in~\cite{lukmeljpikwor,lukmeljwor,meljpikgup,meljsamst2,meljskos}.

\subsection[Right covariant realization of kappa-Minkowski space time]{Right covariant realization of $\boldsymbol{\kappa}$-Minkowski space time}\label{section5.1}

The right covariant realization of $\kappa$-Minkowski space is given by
\begin{gather*}
\hat{x}_\mu=x_\mu -a_\mu x\cdot p, \qquad [\hat{x}_\mu, \hat{x}_\nu]={\rm i}(a_\mu \hat{x}_\nu - a_\nu \hat{x}_\mu).
\end{gather*}
$J_\mu(k,q)$ for the right covariant realization satisfies the following differential equation
\begin{gather*}
\frac{\td J_\mu (tk,q)}{\td t}=k_\mu -a\cdot k J_\mu(tk,q),
\end{gather*}
with initial condition $J_\mu(0,q)=q_\mu$. $J_\mu^{(0)}(tk,q)=tk_\mu +q_\mu$ for $a_\mu=0$. The solution of the above differential equation is
\begin{gather*}
J_\mu(tk,q)=k_\mu \frac{1-\te ^{-ta\cdot k}}{a\cdot k} + q_\mu \te ^{-ta\cdot k}.
\end{gather*}
For $t=1$,
\begin{gather*}
J_\mu (k,q)=k_\mu \frac{1-\te ^{-a\cdot k}}{a\cdot k} + q_\mu \te ^{-a\cdot k}.
\end{gather*}
For $k_\mu=0$, $J_\mu(0,q)=q_\mu$. For $q_\mu=0$,
\begin{gather*}
J_\mu (k,0)=k_\mu \frac{1-\te ^{-a\cdot k}}{a\cdot k}\equiv K_\mu(k).
\end{gather*}
From here it follows that
\begin{gather*}
a\cdot K=1-\te^{-a\cdot k}, \qquad a\cdot k =-\ln (1-a\cdot K(k)), \qquad \te ^{-a\cdot k}= 1-a\cdot K.
\end{gather*}
Hence, the inverse of $K_\mu(k)$ is given by
\begin{gather*}
K_\mu ^{-1}(k)= -k_\mu \frac{\ln(1-a\cdot k)}{a\cdot k}= k_\mu ^W.
\end{gather*}
Furthermore,
\begin{gather*}
 \mathcal{D}_\mu (k,q)=J_\mu \big(K^{-1}(k),q\big)=k_\mu +q_\mu (1-a\cdot k), \\
 \Delta p_\mu=p_\mu \otimes 1 +(1-a\cdot p) \otimes p_\mu.
\end{gather*}

In the Hopf algebroid approach, the twist corresponding to the right covariant realization is given by
\begin{gather*}
\mathcal{F}=\te ^{-{\rm i}p_\beta^W\otimes \hat{x}_\beta} \te ^{{\rm i}p_\alpha\otimes x_\alpha}=\te ^{\mathcal{A}} \te ^{\mathcal{B}},
\end{gather*}
where
\begin{gather*}
 \mathcal{A}=-{\rm i}p_\alpha^W\otimes x_\alpha +{\rm i}a \cdot p^W\otimes x\cdot p, \qquad
 \mathcal{B}={\rm i}p_\alpha\otimes x_\alpha, \qquad
 [\mathcal{A},\mathcal{B}]=\big(a\cdot p^W \otimes 1\big)\mathcal{B}.
\end{gather*}
Using a special case of the BCH formula, we get (see \cite[Appendix C]{jurmeljpik})
\begin{gather*}
\mathcal{F} =\exp\left(\mathcal{A}+\mathcal{B}\left( \frac{a\cdot p^W \otimes 1}{1\otimes 1-\te ^{-a\cdot p^W \otimes 1}}\right)\right)
\end{gather*}
Using the relation
\begin{gather*}
p_\mu^W=-p_\mu\frac{\ln (1-a\cdot p)}{a\cdot p},
\end{gather*}
we get
\begin{gather}\label{Fizsec5.1}
 \mathcal{F}=\te ^{-{\rm i}\ln(1-a\cdot p)\otimes D}, \qquad D=x_\alpha p_\alpha, \qquad \text{and}\qquad
 \Delta p_\mu =p_\mu \otimes 1+(1-a\cdot p)\otimes p_\mu.
\end{gather}
We have shown that the twist in the Hopf algebroid approach corresponding to the right covariant realization of $\kappa$-Minkowski is identical to the Jordanian twist leading to the same right covariant realization and satisfying the cocycle condition in the Hopf algebra sense.

\subsection[Left covariant realization of kappa-Minkowski space time]{Left covariant realization of $\boldsymbol{\kappa}$-Minkowski space time}\label{section5.2}

The left covariant realization of $\kappa$-Minkowski spacetime is given by
\begin{gather*}
\hat{x}_\mu =x_\mu(1+a\cdot p), \qquad [\hat{x}_\mu,\hat{x}_\nu]= {\rm i}(a_\mu \hat{x}_\nu -a_\nu \hat{x}_\mu).
\end{gather*}
$J_\mu(k,q)$ for the left covariant realization satisfies the following differential equation
\begin{gather*}
\frac{\td J_\mu(tk,q)}{\td t}=k_\mu (1+a\cdot J(tk,q)),
\end{gather*}
with initial condition $J_\mu(0,q)=q_\mu$. $J_\mu ^{(0)}(tk,q)=tk_\mu+q_\mu$ for $a_\mu=0$. The solution of the above differential equation is
\begin{gather*}
J_\mu(tk,q)=k_\mu \frac{\te ^{ta\cdot k}-1}{a\cdot k} (1+a\cdot q)+q_\mu.
\end{gather*}
For $t=1$,
\begin{gather*}
J_\mu(k,q)=k_\mu \frac{\te ^{a\cdot k}-1}{a\cdot k}(1+a\cdot q)+q_\mu.
\end{gather*}
For $q_\mu=0$,
\begin{gather*}
J_\mu(k,0)=k_\mu \frac{\te ^{a\cdot k}-1}{a\cdot k} \equiv K_\mu(k).
\end{gather*}
It follows that $a\cdot K=\te ^{a\cdot k}-1$, $a\cdot k=\ln (1+a\cdot K(k))$, and the inverse of $K_\mu(k)$ is given by
\begin{gather*}
K^{-1}_\mu(k)=k_\mu \frac{\ln (1+a\cdot k)}{a\cdot k}=k_\mu ^W.
\end{gather*}
Furthermore,
\begin{gather*}
 \mathcal{D}_\mu (k,q)=J_\mu\big(K^{-1}(k),q\big)=k_\mu (1+a\cdot q)+q_\mu,\qquad
 \Delta p_\mu =p_\mu\otimes (1+a\cdot p)+1\otimes p_\mu.
\end{gather*}

The twist in the Hopf algebroid approach corresponding to the left covariant realization is
\begin{gather*}
\mathcal{F} =\te ^{-{\rm i}p_\beta^W \otimes \hat{x}_\beta} \te ^{{\rm i}p_\alpha \otimes x_\alpha}= \te ^{\mathcal{A}} \te ^{\mathcal{B}},
\end{gather*}
where
\begin{gather*}
 \mathcal{A}= -{\rm i}p_\beta^W \otimes x_\beta -{\rm i}p_\beta^W \otimes x_\beta a\cdot p,\qquad
 \mathcal{B} ={\rm i}p_\alpha \otimes x_\alpha, \\
 [\mathcal{A},\mathcal{B}]= -(\ln(1+a \cdot p) \otimes 1)\mathcal{B} =-\big(a\cdot p^W \otimes 1\big) \mathcal{B}.
\end{gather*}
Using a special case of the BCH formula we get $\mathcal{F}$ (see \cite[Appendix~C]{jurmeljpik})
\begin{gather*}
\mathcal{F}=\exp\left( \mathcal{A} +\mathcal{B}\left(\frac{-a\cdot p^W \otimes 1}{1\otimes 1-\te ^{a\cdot p^W \otimes 1}}\right)\right).
\end{gather*}
Using
\begin{gather*}
p^W_\mu=p_\mu \frac{\ln(1+a\cdot p)}{a\cdot p}, \qquad a\cdot p^W=\ln(1+a\cdot p),
\end{gather*}
we get
\begin{gather*}
 \mathcal{F}=\te ^{-{\rm i}p_\beta^W\otimes x_\beta a\cdot p}= \te ^{-{\rm i}a_\alpha p_\beta^W \otimes L_{\beta\alpha}}, \qquad L_{\beta\alpha}=x_\beta p_\alpha, \qquad
 \Delta p_\mu =p_\mu \otimes (1+a\cdot p)+1\otimes p_\mu.
\end{gather*}
We have shown that the twist in the Hopf algebroid approach corresponding to the left covariant realization of the $\kappa$-Minkowski spacetime is different from the Jordanian twist leading to the same left covariant realization which satisfies the cocycle condition in the Hopf algebra sense, $\mathcal{F}=\exp ( -{\rm i}D\otimes \ln(1+a\cdot p) )$. Although these twists are different, they give the same deformed Hopf algebra. Drinfeld twists $\mathcal{F}=\exp(-{\rm i}\ln (1-a\cdot p)\otimes D)$, \eqref{Fizsec5.1}, and $\mathcal{F}=\exp(-{\rm i}D\otimes \ln(1+ a\cdot p))$ belong to extended Jordanian twists for Lie algebras \cite{kullyamud}. Interpolations between Jordanian twists, right and left covariant realizations of the $\kappa$-Minkowski spacetime were presented in refs. \cite{bormeljpach,meljskos2,meljskos3,meljskos4}.

\subsection[Light like realization of kappa-Minkowski spacetime]{Light like realization of $\boldsymbol{\kappa}$-Minkowski spacetime}\label{section5.3}

The light like realization of the $\kappa$-Minkowski space is defined with $a^2=0$ and
\begin{gather*}
\hat{x}_\mu =x_\mu(1+a\cdot p)-a\cdot x p_\mu =x_\mu +a_\alpha M_{\mu\alpha},
\end{gather*}
satisfying the $\kappa$-Minkowski algebra
\begin{gather*}
 [\hat{x}_\mu,\hat{x}_\nu]= {\rm i}(a_\mu \hat{x}_\nu -a_\nu \hat{x}_\mu), \\
 [M_{\mu\nu}, \hat{x}_\lambda]= -{\rm i}(\hat{x}_\mu \eta_{\nu\lambda} -\hat{x}_\nu \eta_{\mu\lambda})-{\rm i}(a_\mu M_{\nu\lambda}-a_\nu M_{\mu\lambda}), \\
 [M_{\mu\nu},p_\lambda]= -{\rm i}(p_\mu \eta_{\nu\lambda}-p_\nu \eta_{\mu\lambda}).
\end{gather*}

Using \eqref{J-lin},\eqref{mcK-lin}, \eqref{J-qje0}, \eqref{Kinv-lin}, \eqref{D-lin}, one obtains the following expression for~$\mathcal{D}_\mu (k,q)$
\begin{gather*}
\mathcal{D}_\mu (k,q)= k_\mu (1+a\cdot q)q_\mu -a_\mu \frac{k\cdot q}{1+a\cdot k}- \frac{1}{2} a_\mu (a\cdot q) \frac{p^2}{1+a\cdot k},
\end{gather*}
and for the coproduct
\begin{gather}\label{Deltap-light}
\Delta p_\mu =\mathcal{D}_\mu (p\otimes 1,1\otimes p)= \Delta_0 p_\mu +\left( p_\mu a_\alpha -a_\mu \frac{p_\alpha +\frac{1}{2}a_\alpha p^2}{1+a\cdot p}\right) \otimes p_\alpha.
\end{gather}

From the expression for the twist given in equation~\eqref{twist-lin}, it follows that the Drinfeld twist is given by \cite{jurmeljpik,jurmeljpiks,jurmeljsam2}
\begin{gather*}
\mathcal{F}=\exp \left( {\rm i} a_\alpha p_\beta \frac{\ln (1+a\cdot p)}{a\cdot p}\otimes M_{\alpha\beta}\right),
\end{gather*}
which satisfies the cocycle condition in the Hopf algebra sense and the coproduct $\Delta p_\mu =\mathcal{F}\Delta_0 p_\mu \mathcal{F}^{-1}$ is coassociative and coincides with $\Delta p_\mu$ above,~\eqref{Deltap-light}.

\begin{Remark}
If $a^2\neq 0$, the above realization $\hat{x}_\mu =x_\mu(1+a\cdot p) -a\cdot x p_\mu =x_\mu +{\rm i}a_\alpha M_{\mu\alpha}$, leads to the $\kappa$-Snyder algebra \cite{meljsamst1,meljsamst2}
\begin{gather*}
[\hat{x}_\mu,\hat{x}_\nu]={\rm i}(a_\mu \hat{x}_\nu-a_\nu \hat{x}_\mu)+{\rm i}a^2 M_{\mu\nu}.
\end{gather*}
The corresponding twist \eqref{twist-chije0} does not satisfy the cocycle condition, the star product is non associative and the coproduct is non coassociative.
\end{Remark}

\section{Quadratic deformations of quantum phase space}\label{section6}

In \cite{Wess}, deformed Heisenberg algebras were constructed as examples of NC structures and the framework for higher dimensional NC spaces based on quantum groups was studied. Furthermore, quadratic deformations of the Minkowski space from twisted Poincar\'e symmetries were constructed in \cite{lukwor}. The construction was based on the twist
\begin{gather*}
\mathcal{F}=\exp\left( \frac{{\rm i}}{2} \Theta^{\alpha\beta\gamma\delta} M_{\alpha\beta}\wedge M_{\gamma\delta}\right),
\end{gather*}
where $M_{\alpha\beta}$ are Lorentz generators and the $r$-matrix is given with $r=\frac{1}{2}\Theta_{\mu\nu\rho\sigma} M_{\mu\nu}\wedge M_{\rho\sigma}$. Quadratic deformations of the Galilei group and the Newton equation for classical space were considered in~\cite{daskwal}.

Generally, quadratic algebras, i.e., quadratic deformations, can be defined with the following commutation relations
\begin{gather*}
[\hat{x}_\mu,\hat{x}_\nu]= \Theta^{\mu\nu\gamma\delta} \hat{x}_\gamma \hat{x}_\delta,
\end{gather*}
where all Jacobi relations have to be satisfied and multiplication in the enveloping algebra $\mathcal{U}(\hat{x})$ is associative. In this case the general realization of the NC coordinates is
\begin{gather*}
\hat{x}_\mu= x_\alpha\varphi_{\alpha\mu} ({\rm i}L_{\gamma\delta}),
\end{gather*}
where $L_{\gamma\delta}=x_\gamma p_\delta$ generate the $\mathfrak{gl}(n)$ algebra
\begin{gather*}
[L_{\mu\nu}, L_{\rho\sigma}]={\rm i}(\eta_{\mu\sigma}L_{\rho\nu} -\eta_{\rho\nu}L_{\mu\sigma}).
\end{gather*}
In the lowest order in $K_{\mu\nu\gamma\delta}$ we get
\begin{gather*}
\hat{x}_\mu=x_\mu +{\rm i}x_\alpha L_{\beta\gamma} K_{\mu\gamma\beta\alpha} +O\big(K^2\big) =x_\mu +{\rm i}K_{\mu\gamma\beta\alpha} x_\alpha x_\beta p_\gamma +O\big(K^2\big),
\end{gather*}
with $K_{\mu\nu\gamma\delta}-K_{\nu\mu\gamma\delta}=\Theta_{\mu\nu\gamma\delta}$.

We point out that Sections~\ref{section3} and~\ref{section4} cannot be applied to these quadratic deformations. The construction of such quadratic algebras can be performed using twist operators that, besides Lorentz generators, include dilatation operators $D_\mu$ and $D$
\begin{gather*}
D=\sum_{\mu} D_\mu=\sum_{\mu} x_\mu p_\mu,\qquad
D_\mu = x_\mu p_\mu \quad \text{(no summation)},
\end{gather*}
and more generally $L_{\mu\nu}=x_\mu p_\nu$ generating the $\mathfrak{gl}(n)$ algebra. Here we consider a simple case.

\subsection{Quadratic deformations of Minkowski space from dilatation}\label{section6.1}

We consider the twist
\begin{gather}\label{twistsec61}
\mathcal{F}=\exp\bigg( \sum_{\alpha,\beta} a_{\alpha\beta} D_\alpha \otimes D_\beta\bigg),\qquad a_{\beta\alpha}=-a_{\alpha\beta},
\end{gather}
where
\begin{gather*}
 [D_\alpha,D_\beta] = 0, \qquad
 [D_\alpha, p_\beta] = {\rm i}p_\alpha \eta_{\alpha\beta} \quad \text{(no summation)}, \qquad
 [D_\alpha, x_\beta] = -{\rm i}x_\alpha \eta_{\alpha\beta},
\end{gather*}
and $p_\alpha$ are momenta.

The action of $D_\alpha$ in undeformed quantum phase space is defined as
\begin{gather*}
D_\alpha \bt f(x) =-{\rm i}x_\alpha \frac{\partial f(x)}{\partial x_\alpha}.
\end{gather*}
The deformed quantum phase space is defined with
\begin{gather*}
 \hat{x}_\alpha \hat{x}_\beta =q_{\alpha\beta} \hat{x}_\beta \hat{x}_\alpha, \qquad
 q_{\alpha\beta} =\exp ( a_{\alpha\beta} -a_{\beta\alpha} )= \exp (2a_{\alpha\beta}), \qquad
 [D_\alpha, \hat{x}_\beta]=-{\rm i}\hat{x}_\alpha \eta_{\alpha\beta}, \\
 p_\alpha \hat{x}_\beta -\te ^{a_{\beta\alpha}} \hat{x}_\beta p_\alpha =-{\rm i}\eta_{\alpha\beta} \exp\bigg( \sum_{\gamma} {\rm i}a_{\beta\gamma} D_\gamma \bigg) , \qquad c_{\alpha\beta} =\te ^{a_{\alpha\beta}}.
\end{gather*}
The realization for $\hat{x}_\alpha$ is given by
\begin{gather*}
\begin{split}
& \hat{x}_\alpha =x_\alpha \exp\bigg( \sum_{\beta} {\rm i}a_{\alpha\beta} D_\beta \bigg)= x_\alpha \phi_\alpha, \qquad
 \phi_\alpha =\exp\bigg( \sum_{\beta} {\rm i}a_{\alpha\beta} D_\beta \bigg), \\
& \phi_\alpha \bt 1= 1, \qquad \hat{x}_\alpha \bt 1=x_\alpha.
\end{split}
\end{gather*}
Note that
\begin{gather*}
 x_\alpha \phi_\alpha =\exp(-a_{\alpha\alpha}\phi_\beta x_\alpha), \qquad
 x_\alpha \phi_\beta =\exp (-a_{\beta\alpha} \phi_\beta x_\alpha), \\
 x_\alpha f({\rm i}D_\alpha)= f({\rm i}D_\alpha -1)x_\alpha, \qquad
 p_\alpha f({\rm i}D_\alpha)=f({\rm i}D_\alpha +1)p_\alpha.
\end{gather*}
The coproducts are given by
\begin{gather*}
 \Delta D_\alpha =\mathcal{F} \Delta_0 D_\alpha \mathcal{F}^{-1}= D_\alpha\otimes 1+1\otimes D_\alpha=\Delta_0 D_\alpha, \qquad
 \Delta \phi_\alpha = \phi_\alpha \otimes \phi_\alpha \!\quad \text{(no summation)}, \\
 \Delta p_\alpha= \mathcal{F} \Delta_0 p_\alpha \mathcal{F}^{-1}= p_\alpha \otimes \phi_\alpha+\tilde{\phi}_\alpha \otimes p_\alpha, \\
 \tilde{\phi}_\alpha =\exp\bigg( \sum_{\alpha} {\rm i}a_{\beta\alpha} D_\beta \bigg), \qquad
 \phi_\alpha =\exp\bigg( \sum_{\beta} {\rm i}a_{\alpha\beta} D_\beta \bigg), \\
 \Delta \tilde{\phi}_\alpha =\tilde{\phi}_\alpha \otimes \tilde{\phi}_\alpha, \qquad
 \phi_\alpha \hat{x}_\beta =\te ^{a_{\alpha\beta}} \hat{x}_{\beta} \phi_\alpha, \qquad
 [\phi_\alpha ,\hat{x}_\beta]=\big( \te ^{a_{\alpha\beta}}-1\big) \hat{x}_\beta \phi_\alpha.
\end{gather*}
The twist $\mathcal{F}$ given in equation~\eqref{twistsec61} is abelian and satisfies the cocycle condition. The star product is associative
\begin{gather*}
 x_\alpha *x_\beta =\hat{x}_\alpha \hat{x}_\beta \bt 1= \hat{x}_\alpha \bt x_\beta = \te ^{a_{\alpha\beta}}x_\alpha x_\beta, \\
 x_\beta *x_\alpha =\te ^{a_{\beta\alpha}} x_\alpha x_\beta, \qquad
 x_\alpha *x_\beta = q_{\alpha\beta}x_\beta *x_\alpha, \\
 \hat{x}_\alpha =m\mathcal{F}^{-1} (\bt \otimes 1)(x_\alpha \otimes 1)= x_\alpha \exp\bigg( \sum_{\beta} a_{\alpha\beta} D_\beta\bigg), \\
 (f*g)(x)=m\mathcal{F}^{-1}(\bt \otimes \bt)(f(x)\otimes g(x)).
\end{gather*}

For $a_{\beta\alpha}=-a_{\alpha\beta}$, $q_{\alpha\beta}=\exp(2a_{\alpha\beta})=(c_{\alpha\beta})^2$, $c_{\alpha\alpha}=q_{\alpha\alpha}=1$, $\tilde{\phi}_\alpha=(\phi_\alpha)^{-1}$.

The $\ct$ action is defined with
\begin{gather*}
 \hat{f}\ct \hat{g}=\hat{f}\hat{g}, \qquad
 \hat{x}_\alpha \hat{f}\ct 1=\hat{x}_\alpha \hat{f}, \qquad
 \hat{x}_\alpha \hat{f}=\big(O_\alpha \ct \hat{f}\big)\hat{x}_\alpha, \qquad
 \hat{f}\hat{x}_\alpha =\hat{x}_\alpha \big(O_\alpha^{-1}\ct \hat{f}\big), \\
 O_\alpha =\phi_\alpha \big(\tilde{\phi}_\alpha\big)^{-1}, \qquad
 O_\alpha \ct 1=1, \qquad
 \phi_\alpha \ct 1=1, \qquad
 p_\alpha \ct 1 =0, \qquad
 p_\alpha \ct \hat{x}_\beta =-{\rm i}\eta_{\alpha\beta}.
\end{gather*}

Let us define
\begin{gather*}
\hat{y}_\alpha = m\tilde{\mathcal{F}}^{-1} (\bt \otimes 1)(x_\alpha \otimes 1)
 = x_\alpha \tilde{\phi}_\alpha =\hat{x}_\alpha (\phi_\alpha)^{-1}\tilde{\phi}_\alpha
 = \hat{x}_\alpha (O_\alpha)^{-1},
\end{gather*}
where $\tilde{\mathcal{F}}=\mathcal{F}^{\rm op}$, $\mathcal{F}^{\rm op}=\exp\big( \sum_{\alpha,\beta} a_{\alpha\beta}D_\beta \otimes D_\alpha \big)$,
\begin{gather*}
 \hat{y}_\alpha \bt 1=x_\alpha,\qquad
 \hat{y}_\alpha \ct \hat{f}=\hat{f}\hat{x}_\alpha.
\end{gather*}
For $a_{\beta\alpha}=-a_{\alpha\beta}$, $O_\alpha=(\phi_\alpha)^2$.
\begin{gather*}
 [\hat{x}_\alpha,\hat{y}_\beta]=0 \quad \forall \alpha,\beta,\qquad
 \hat{y}_\alpha\hat{y}_\beta =\exp(-2a_{\alpha\beta})\hat{y}_\beta \hat{y}_\alpha.
\end{gather*}
The special case where $q_{\alpha \beta} = q$ for $\alpha > \beta$ and $q_{\alpha \beta} = q^{-1}$ for $\alpha < \beta$ was studied in \cite{koornwinder,koorsw}.
\begin{Remark}
For $a_{\beta\alpha}=a_{\alpha\beta}$ it follows that $\mathcal{F}=\mathcal{F}^{\rm op}$, $\hat{x}_\mu=\hat{y}_\mu$, $[\hat{x}_\mu,\hat{x}_\nu]=0$, $\tilde{\phi}_\mu=\phi_\mu$ and $(f*g)(x)=(g*f)(x)$, but $c_{\alpha\alpha}\neq 1$ and $p_\alpha \hat{x}_\beta -\te ^{a_{\alpha\beta}} \hat{x}_\beta p_\alpha =-{\rm i}\eta_{\alpha\beta} \exp\big(\sum_\gamma {\rm i}a_{\beta\gamma}D_\gamma\big)$. For the case of one dimension, $n=1$, see~\cite{bardmelj, Wess}. Applications to the Fock space representation and the Calogero model in one dimension were considered in~\cite{bardmelj}.
\end{Remark}

\section{Generalization of Yang and triply special relativity models}\label{section7}

In Sections~\ref{section2}--\ref{section5}, we have considered quantum deformed phase spaces (for example the $\Theta$ canonical space, Lie algebra type spaces, the Snyder space) in which defomation parameters are proportional to the minimal length~$l$. In Section~\ref{section6}, related to quadratic deformations of quantum phase space, deformation parameters are dimensionless. There are also deformed quantum phase spaces in which deformation parameters depend on two physical quantities, the minimal length and the cosmological radius~$R$. These models are generated with NC coordinates~$\hat{x}_\mu$ and NC momenta~$\hat{p}_\mu$. A large class of such deformed quantum phase spaces can be described with algebras containing~2 Snyder algebras as subalgebras, with the same Lorentz algebra generated with~$M_{\mu\nu}$. They are defined as
\begin{gather}
[\hat{x}_\mu,\hat{x}_\nu]={\rm i}\beta^2M_{\mu\nu},\label{sec7-1}\\
[\hat{p}_\mu,\hat{p}_\nu]= {\rm i}\alpha^2M_{\mu\nu},\\
[M_{\mu\nu}, \hat{x}_\lambda]={\rm i}(\eta_{\mu\lambda} \hat{x}_\nu -\eta_{\nu\lambda} \hat{x}_\mu),\\
[M_{\mu\nu},\hat{p}_\lambda]= {\rm i}(\eta_{\mu\lambda}\hat{p}_\nu -\eta_{\nu\lambda}\hat{p}_\mu),\\
[M_{\mu\nu},M_{\rho\sigma}]= {\rm i}(\eta_{\mu\rho}M_{\nu\sigma} -\eta_{\mu\sigma} M_{\nu\rho} -\eta_{\nu\rho}M_{\mu\sigma} +\eta_{\nu\sigma}M_{\mu\rho}),\\
[\hat{x}_\mu,\hat{p}_\nu]= {\rm i}g_{\mu\nu},\\
[M_{\mu\nu},g_{\rho\sigma}]= {\rm i}(\eta_{\mu\sigma}g_{\rho\nu}-\eta_{\nu\sigma} g_{\rho\mu} +\eta_{\mu\rho} g_{\nu\sigma} -\eta_{\nu\rho}g_{\mu\sigma}),\\
[g_{\mu\nu},\hat{x}_\lambda]-[g_{\lambda\nu},\hat{x}_\mu] ={\rm i}\beta^2 (\eta_{\mu\nu} \hat{p}_\lambda -\eta_{\lambda\nu} \hat{p}_\mu),\\
[g_{\nu\mu},\hat{p}_\lambda]-[g_{\nu\lambda},\hat{p}_\mu] ={\rm i}\alpha^2(\eta_{\lambda\nu} \hat{x}_\mu -\eta_{\mu\nu} \hat{x}_\lambda)=-{\rm i}\alpha^2(\eta_{\mu\nu} \hat{x}_\lambda -\eta_{\lambda\nu} \hat{x}_\mu),\\
[g_{\mu\nu},g_{\rho\sigma}]={\rm i}\big([[g_{\mu\nu},\hat{p}_\sigma],\hat{x}_\rho]- [[g_{\mu\nu},\hat{x}_\rho],\hat{p}_\sigma]\big).\label{sec7-2}
\end{gather}
These algebras are Born dual, $\hat{x}_\mu \leftrightarrow \hat{p}_\mu$, $M_{\mu\nu} \leftrightarrow M_{\nu\mu}$, $g_{\mu\nu} \leftrightarrow -g_{\nu\mu}$, $\alpha \leftrightarrow \beta$.

Hermitian realizations of $\hat{x}_\mu$, $\hat{p}_\mu$, $M_{\mu\nu}$ and $g_{\mu\nu}$ can be written as
\begin{gather*}
 \hat{x}_\mu =\tfrac{1}{2}\big(x_\mu F+F^\dagger x_\mu +p_\mu G+G^\dagger p_\mu\big), \\
\hat{p}_\mu =\tfrac{1}{2}\big(p_\mu H+H^\dagger p_\mu +x_\mu K+K^\dagger x_\mu\big), \\
M_{\mu\nu} =x_\mu p_\nu -x_\nu p_\mu, \\
g_{\mu\nu} =\eta_{\mu\nu}h_0 +\alpha^2 x_\mu x_\nu h_1 +h_1^\dagger \alpha^2 x_\mu x_\nu +\beta^2 p_\mu p_\nu h_2 +h_2^\dagger \beta^2 p_\mu p_\nu+\alpha \beta (x_\mu p_\nu +p_\nu x_\mu )h_3 \\
\hphantom{g_{\mu\nu} =}{}
+ h_3^\dagger \alpha \beta (x_\mu p_\nu +p_\nu x_\mu ) +\alpha \beta (x_\nu p_\mu +p_\mu x_\nu)h_4 +h_4^\dagger \alpha \beta (x_\nu p_\mu +p_\mu x_\nu) , \\
g_{\mu\nu}-g_{\nu\mu} =2M_{\mu\nu} \alpha\beta \big(h_3+ h_3^\dagger -h_4- h_4^\dagger\big),
\end{gather*}
where $F$, $G$, $H$, $K$, $h_0$, $h_1$, $h_2$, $h_3$, $h_4$ are Lorentz invariants depending on $x^2$, $x\cdot p$, $p^2$. For the Yang model, $g_{\mu \nu} = \eta_{\mu \nu} h_0$.

Yang quantum phase spaces \cite{guohuangwu, yang} and tripy special relativity models \cite{kowgliksmol,mig2,mig3} are special cases of the above deformed quantum phase spaces. Realizations of these models are more difficult to construct and will be presented elsewhere. Spinorial Snyder and Yang models from super algebras and NC quantum super spaces have recently been constructed \cite{zadnja}. Similarly, generalized spinorial models, super algebras and quantum super spaces can be constructed by extending the above deformed quantum phase spaces \eqref{sec7-1}--\eqref{sec7-2}.

\subsection*{Acknowledgement}

SM thanks S.~Mignemi for useful comments.

\pdfbookmark[1]{References}{ref}
\LastPageEnding

\end{document}